\documentclass[12pt,amsmath,superscriptaddress,nofootinbib]{revtex4}

\usepackage{graphicx}
\usepackage{enumitem}
\usepackage{color}
\usepackage{hyperref}

\newcommand{\beq}{\begin{eqnarray}}
 \newcommand{\eeq}{\end{eqnarray}}
\newcommand{\be}{\begin{equation}}
 \newcommand{\ee}{\end{equation}}

\def\fun#1#2{\lower3.6pt\vbox{\baselineskip0pt\lineskip.9pt
\ialign{$\mathsurround=0pt#1\hfil ##\hfil$\crcr#2\crcr\sim\crcr}}}
\newcommand{\tr}{\operatorname{tr }}

\newcommand{{\SD}}{\rm SD}

\newcommand{{\Mc}}{\mathcal{M}}

\newcommand{\vep}{\mbox{\boldmath${\rm p}$}}

\newcommand{\veR}{\mbox{\boldmath${\rm R}$}}

\newcommand{\llan}{\langle\langle}
\newcommand{\rran}{\rangle\rangle}
\newcommand{\lan}{\langle}
\newcommand{\ran}{\rangle}

\begin{document}

\title{Dynamics of the Polyakov loops in QCD}

\author{R.A.Abramchuk}
\affiliation{Moscow Institute of Physics and Technology, 9, Institutskii per., Dolgoprudny, Moscow Region, 141700, Russia}

\author{Z.V.Khaidukov}
\affiliation{Moscow Institute of Physics and Technology, 9, Institutskii per., Dolgoprudny, Moscow Region, 141700, Russia}
\affiliation{Institute for Theoretical and Experimental Physics, B. Cheremushkinskaya 25, Moscow, 117259, Russia}

\author{Yu.A.Simonov}
\email{simonov@itep.ru}
\affiliation{Institute for Theoretical and Experimental Physics, B. Cheremushkinskaya 25, Moscow, 117259, Russia}

\date{\today}

\begin{abstract}
A non-perturbative method of Field Correlators is applied to calculate the Polyakov loop dependence on temperature, $L(T)$,  in the  2+1  flavor QCD with small quark masses, so the only relevant scale is the color-electric string tension $\sigma(T)$.
Polyakov loop in the temperature range 100 MeV $\lesssim T\lesssim$ 400 MeV is expressed via the heavy-light meson mass that decreases with $T$ as $\sqrt{\sigma (T)}$.
The latter is deduced from a gradually vanishing chiral condensate. 
The resulting $L(T)$ is in good agreement with recent lattice data.
\end{abstract}

\maketitle

\section{Introduction}

The Polyakov loop $L(T)$ has a long history and ever increasing role in understanding the QCD dynamics at nonzero temperature. 

$L_i(T)$  ($i=f,\, adj$ stand for fundamental and adjoint representations) may be associated with the renormalized free energy of a static charge $L(T) = \exp (-F_Q/T)$.
The first studies \cite{1,2,3,4} that exploited this association discovered a dynamical mechanism that may explain the Polyakov loop role in the confined and deconfined phases of SU(3)-gluodynamics, where $L(T)$ can be considered as an order parameter. 

In the case of gluodynamics, $L_f(T)$ is associated with the free energy of a static quark \(F_Q\). 
In the confined phase, the free energy of a static quark isolated from an antiquark is infinite due to the presence of an infinitely long string; whereas the free energy is finite in the deconfined phase, implying that \(L_f(T)=0\) at \(T<T_c\).
This result together with Casimir scaling studies was obtained in \cite{5}.
  
The studies \cite{6,7,8,9,10,11,12} of the $n_f =2+1$ QCD revealed a completely different behavior of $L_f(T)$. 
In the presence of dynamical quarks an isolated static quark adjoins a dynamical antiquark. 
As a result, the free energy $F_Q$ of the system is finite, and $L_f(T)$ is nonzero even in the confined phase.  
Therefore, the temperature dependence of the Polyakov loop is continuous. 
Moreover, $L_i (T)$ may be associated with the $Q\bar Q$ free energy and the Debye screening mass \cite{6} at temperatures far above the deconfinement temperature \(T_c\sim 140\) MeV.
  
Renormalized Polyakov loop was calculated repeatedly on lattice in $n_f =2+1$ QCD with physical quark masses \cite{7,8,9,10,11,12}. 
In these studies continuous curves $L_{\rm ren}(T)$ strongly depend on a lattice quark character and an applied renormalization procedure.
  
In spite of numerous studies, two main questions concerning Polyakov loops remain unanswered:
\begin{enumerate}
    \item What is the theoretical dynamics behind the Polyakov loop, and how to calculate \(L_i(T)\) starting with the QCD Lagrangian? 
    \item What is the explicit role of $L_i(T)$ in the expressions for the thermodynamic potentials in gluodynamics and QCD?
\end{enumerate}
  
Though the lattice data on $L(T)$ are obtained numerically from the first principles, the data do not reveal the dynamical mechanisms for perturbative and non-perturbative interactions generating $L(T)$.

Partial answers to both questions have been formulated in the framework of a non-perturbative approach called the Field Correlator Method (FCM) \cite{13,14}. 
As for QCD thermodynamics within the FCM, and for the review, see \cite{15,16,17,18,19,20} and \cite{21}, respectively.
Recently the formalism was generalized to include the Color Magnetic  Confinement (CMC) at $T>T_c$ \cite{22,23,24,25,26}. 
  
Regarding the second question, within the FCM $L_i(T)$ enters the resulting expressions for thermodynamic quantities in the deconfined phase.
For the pressure of strongly interacting quark-gluon plasma (QGP), the following expression was obtained \cite{25}
  \be P=\sum_f P^{(f)}_q + P_{gl},\label{1.1}\ee
  \be P_{gl} = \frac{N_c^2-1}{\sqrt{4\pi}} \int^\infty_0 \frac{ds}{s^{3/2}} G_3 (s) \sum_{n=0}^{\infty} e^{- \frac{n^2}{4T^2s}} L_{\rm adj}^{(n)},\label{1.2}\ee
  \be P_q^{(f)} = \frac{4N_c}{\sqrt{ 4\pi}} \int^\infty_0 \frac{ds}{s^{3/2}} e^{-m^2_f s} S_3 (s) \sum_{n= 1}^{\infty}(-)^{n+1} e^{- \frac{n^2}{4T^2s}}\cosh \left( \frac{\mu n}{T}\right)  L_{f}^{(n)},\label{1.3}\ee
where $G_3(s)$ and $S_3(s)$ are 3d two-point Greens' functions of gluons and quarks (with the spinor or tensor indices contracted), respectively.
Casimir scaling provides \(L_{adj}=(L_f)^{9/4}\).

The Greens' functions depend on CMC screening masses $m_D\approx  const \sqrt{\sigma_s}$, where $\sigma_s$ is spatial, or color-magnetic (CM), string tension.

At $T<1$ GeV the following approximation for a \(n\)-times wound Polyakov loop is applicable \cite{18}
\be L_i^{(n)}  \approx (L_i)^n = L_i^n \label{1.4}\ee
  
According to \eqref{1.1}-\eqref{1.4}, to a large extent, the Polyakov loop defines the QGP (and gluon plasma) thermodynamics.
For example, in the gluodynamics at $T>T_c$, the behavior of $L_{adj} (T)$ defines the spectacular shoulder in the $T$ dependence of $\frac{I(T)}{T^2T^2_c}$ \cite{22}, where  $I(T)$ is the trace anomaly.

However, $L(T)$ is irrelevant to thermodynamics of the confined QCD --- all quarks and gluons in this  region are bound inside  hadrons.
In the confined QCD, the pressure dominantly has the form of hadron resonance gas (HRG) pressure.

The importance of the Polyakov loop is also recognized in other approaches.  
An effective formalism for the QGP thermodynamics description is developed, for instance, in the Polyakov--Nambu--Jona--Lasinio model \cite{27,28} and in the Polyakov-quark-meson model \cite{28*, 28**}. 

An answer to the first question was found partially within the FCM framework in \cite{18,19}.
$L_i(T)$ at \(T>T_c\) was expressed via a non-perturbative  $(np)$ interaction $V_1(r,T)$ produced by a field correlator $D_1^E (z)$. 
The perturbative part of $V_1$ coincides with the standard color Coulomb interaction. 
The long distance $ np $ part of $V_1$ produces $L_i(T)$ as
  \be L_i (T) =\exp \left( -c_i \frac{V_1(\infty,T)}{2T}\right), ~~ c_f=1, ~ c_{adj} =\frac94.\label{1.5}\ee

As shown in \cite{18,19} and discussed below, $V_1$ can be obtained from the correlator $D_1^E$ that is obtained from the 1-gluon gluelump propagator, known analytically at $T=0$ \cite{29} and on the lattice \cite{30}.

In the present paper, we demonstrate the connection between the Polyakov loop in QCD and the heavy-light mass at 100 MeV \(\lesssim T\lesssim\) 400 MeV. 
Then we extract the heavy-light mass dependence on temperature from lattice data on the chiral condensate \cite{52,53} and compare the obtained \(L_f(T)\) with direct lattice computations in \(n_f=2+1\) QCD \cite{56,59}.
The good agreement validates our assumption about the theoretical dynamics behind the Polyakov loop in QCD within the temperature range.

\section{Calculation of the $V_1$ potential}

Let us first consider zero-temperature for simplicity. 
Later in the section, we will introduce nonzero temperature in a standard way by means of temporal dimension compactification.

\subsection{Zero-temperature potentials}

To understand how the $q\bar q$ or $gg$ interaction is created in the non-perturbative euclidean vacuum, we use the path integral representation of the $q\bar q$  ($gg$) Green's function \cite{31} in the following form \cite{14}

\be G_{q\bar q, gg} (x,y) = \int d\Gamma_{q\bar q, gg} W (C_{xy}), \label{5.1}\ee
where $d\Gamma$ includes the integration over all paths of $q$ and $\bar q$ ($g$ and $g$) connecting points $x$ and $y$.

Each pair of these paths forms a closed loop $C$ that defines a Wilson loop $ W(C)$
\be W(C) = \lan \tr P\exp (ig \int_{C} A_\mu dz_\mu)\ran, \label{5.2}\ee
where $P$ is the ordering operator. 

The non-abelian Stokes theorem and the cluster expansion \cite{32} yield

\be W(C) =\tr\exp \sum^\infty_{n=1} \frac{(ig)^n}{n!} \int d\sigma (1) d\sigma(2)... d\sigma(n) \llan F (1) ... F(n),\rran\label{5.3}\ee
where the surface elements and the field operators enter within gauge-invariant combinations by means of the parallel transporter $\Phi(a,b) = \exp (ig \int^a_b A_\mu (u) d u_\mu $ as 
\[d\sigma (i) F(i) \equiv d\sigma_{\mu\nu}(u_i)\,\Phi(x_0, u_i) F_{\mu\nu}(u_i) \Phi(u_i, x_0)\] 

Truncation of the sum in \eqref{5.3} at the first non-zero (\(n=2\)) term, the ``Gaussian approximation,'' and the choice of the minimal area surface with the boundary \(C\) provides 4\% accuracy, as argued in \cite{35} on the grounds of the Casimir scaling \cite{36}.

This approximation yields an instantaneous interaction potential \(\hat V (\veR(\tau), \tau)\) between $q$ and $\bar q$ ($g$ and $g$) at time \(\tau\) when the distance between the interacting particles on a trajectory loop \(C_{xy}\) is \(\veR(\tau)\)
\begin{align}
	W_2(C_{xy}) &= \tr \exp \left( - \frac{g^2}{2} \int \int d\sigma (1) d\sigma (2) \llan F(1)  F(2)\rran \right) \\
	&=\exp\left(-\int^{x_4-y_4}_0 \hat V (\veR(\tau), \tau) d\tau\right).\label{5.4}
\end{align}

In what follows, we calculate the spin-independent potential $\hat V(\veR(\tau), \tau)$ following the basic papers \cite{13,14} (for the spin-dependent part, see \cite{37}).

Starting with the standard definition of the quadratic correlator \cite{13} 
\begin{align}
	D_{\mu\nu,\lambda\sigma} (x,0) &= \frac{g^2}{N_c} \lan \tr F_{\mu\nu} (x)  \Phi (x,0) F_{\lambda\sigma} (0)\ran\nonumber\\
	&= (\delta_{\mu\lambda} \delta_{\nu\sigma} -\delta_{\mu\sigma} \delta_{\nu\lambda})  D(x)\nonumber \\
	&\quad +\frac12 \left[ \frac{\partial}{\partial x_\mu} (x_{\lambda} \delta_{\nu\sigma}-x_\sigma \delta_{\nu\lambda} ) +(\mu\lambda \to \nu\sigma)\right]D_1(x).\label{1}
\end{align}
we express $\hat V$ via $D$ and $D_1$ using (\ref{5.4}) 
\begin{align}
    \int \hat Vd\tau &= \frac12 \int D_{14,14} (u-v) d^2u d^2v \nonumber \\
    &=\frac12 \int d\left( \frac{ u_4+v_4 }{2}\right)  d(u_4-v_4) d\left( \frac{ u_1+v_1 }{2}\right)  d(u_1-v_1)D_{14,14} (u-v) \nonumber\\
    &=2 \int dt_4 \int^\infty_0 d\nu \int^R_0 d\eta(R-\eta) \left(D(\sqrt{\nu^2+\eta^2}) + \frac12 \frac{d}{d\eta} (\eta D_1 (\sqrt{\nu^2+\eta^2}))\right) \label{1.a} 
\end{align}
where $t_4 =\frac{u_4+v_4}{2}, ~\nu=|u_4-v_4|, ~\eta=|u_1-v_1|$, and finally obtain
\begin{gather}
    V =V_D (R) + V_1(R) \\
    V_D =2 \int^\infty_0 d\nu \int^R_0 d\eta (R-\eta) D(\sqrt{\nu^2+\eta^2}) \label{1.b} \\
    V_1(r) = \int^r_0 \lambda d\lambda \int^\infty_0 d\tau D^E_1 (\sqrt{\lambda^2 +\tau^2}) \label{2}
\end{gather}
 
As a new step, we express $D_1(x)$ via one-gluon gluelump propagator to the lowest order in the background perturbation theory.

To do this, we extract the part with derivatives from $F_{\mu\nu} (x) = \partial_\mu A_\nu -\partial_\nu A_\mu - ig [A_\mu, A_\nu]$ 
  \be  D_{1~ \mu\nu, \lambda\sigma}^{(0)} (x,y) = 
  \frac{g^2}{2N_c^2}
   \left\{ \frac{\partial}{\partial x_\mu} \frac{\partial}{\partial y_\lambda} \lan  \tr A_\nu (x) \Phi (x,y) A_\sigma (y) \ran +{\rm perm}\right\}. \label{2*}\ee
The structure in the angular brackets is the gluelump propagator --- a gauge invariant combination of the gluon propagator augmented by the parallel transporter.
We denote the gluelump propagator as
$G^{(1g)}_{\nu\sigma} = \delta_{\nu\sigma}  G ^{(1g)} (z)$
assuming the tensor structure.

For $\mu=\lambda=4$ \eqref{2*} yields
\be  \frac{\partial}{\partial x_4} \frac{\partial}{\partial y_4} G_{\nu\sigma}^{(1g)} (x-y)=\frac{\partial}{\partial x_4} (x_4-y_4) D_1(x-y) \delta_{ \nu\sigma}+{\rm perm}. \label{2**}\ee 
so $D_1$ is related to \(G^{(1g)}\)
 \be D_1(x) = - \frac{2g^2}{N^2_c} \frac{dG^{(1g)}}{dx^2}.\label{3}\ee
  
\(G^{(1g)}\) including the gluelump spectrum was found analytically \cite{29} in agreement with the lattice data \cite{30} within the errors estimated.

Inserting (\ref{3}) into (\ref{2}) we obtain the relation between $V_1$ and $G^{(1g)}$
 \be  V_1(r) = -\frac{g^2}{N_c^2} \int^\infty_0 d\tau (G^{(1g)} (\sqrt{r^2 +\tau^2})-G^{(1g)} ( \tau  )). \label{4}\ee

The long-distance potential is readily expressed via the gluelump propagator using \eqref{1.5}, since $G^{(1g)} (x\to \infty )\to 0$
\be  V_1(\infty) = \frac{g^2}{N_c^2} \int^\infty_0 d\tau  G^{(1g)} ( \tau  ).  \label{5}\ee

To numerically estimate \(V_1(\infty)\) we utilize the large \(x\) asymptotic of $G^{(1g)} (x)$ \cite{29}
\be G_{as}^{1g}(x) = \frac{6N^2_c \sigma_f}{4\pi} \exp (-M_{GP} x).\label{7}\ee
A gluelump acquires mass  $M_{GP} \sim 1$ GeV due to confinement. 
The approximation yields
\be D_1^{ {as}} (x) = \frac{6\alpha_s \sigma_f M_{GP}}{x} \exp (-M_{GP} x).\label{6}\ee
At $\alpha_s = 0.3, ~~ M_{GP}=1.4$ GeV \cite{29,30} 
\be V_1^{(as)}(\infty)= \frac{6\alpha_s \sigma_f}{ M_{GP}} \approx  0.23 ~{\rm GeV}. \label{8}\ee

\subsection{Renormalization of the zero-temperature \(V_1\) potential}

We extract the small distance behavior of the gluelump propagator following Appendix 1 of \cite{42}
\be G^{(1g)}(x) \approx \frac{N_c (N_c^2 -1)}{4\pi^2 x^2} \left(1- \frac{\omega^2 x^2}{4} + ...\right), \label{24**}\ee
where $\omega^2 = \frac{g^2}{12N_c} \lan \tr F^2\ran x^2$. 
It yields
\be D_1 (x) \approx \frac{4C_2 \alpha_s}{\pi} \left\{ \frac{1}{x^4} + \frac{\pi^2 G_2}{24 N_c} +...\right\},\label{24***}\ee
where $G_2=\frac{\alpha_s}{\pi} \lan F^a_{\mu\nu} F^a_{\mu\nu} \ran $, so that $D(0) + D_1(0) = \frac{\pi^2}{18} G_2$ is in agreement with analytic and lattice calculations \cite{38,39,40,41}.

We deduce from \eqref{24***} that the singular part of $D_1$ corresponds to the one-gluon-exchange (OGE), or color Coulomb part, and  the non-singular part refers to the renormalized and finite at small distance \(np\) part of \(V_1\).

To properly renormalize the perturbative part of \(V_1\), we examine the potential considering three approximations of $G^{(1g)} (x)$:
\begin{enumerate}[label=\alph*), ref=\alph*)]
    \item \label{itm:as} asymptotic  form at large $x$
    \item \label{itm:el} ``elementary'' gluelump approximation with a definite gluelump mass
    \item \label{itm:fr} free gluon approximation
\end{enumerate}

In case \ref{itm:as} using \eqref{7}, \eqref{6} we obtain 
\be V_1^{ (as)}(r) = 6\alpha_s \sigma_f\int^\infty_0  d\nu (e^{-\nu M_{GP}}- e^{-\sqrt{r^2+\nu^2}M_{GP}}), ~~ V_1^{(as)} (0) =0,\label{13}\ee

In case \ref{itm:el} we can examine the small $x$ behavior of $G^{(1g)}(x)$.
Corresponding Greens' function $G^{(1g)}_m$ has the form of a massive gluon Greens' function
\be  G^{(1g)}_m(x) = \frac{N_c(N_c^2-1)}{4\pi^2}\frac{m}{|x|} K_1 (m|x|), ~~ m\sim M_{GP},\label{9}\ee 
where $K_1$ is the modified Bessel function. 
It yields
\be D_1^{ (m)} (x) = \frac{ m^2\alpha_s (N_c^2-1)}{ \pi x^2 N_c} K_2 (m|x|),\label{10}\ee
thus \(V_1\) is divergent
\be V_1^{(m) } (r)  = \frac{  \alpha_s (N_c^2-1)}{ \pi  N_c}   \int^\infty_0
d\nu \left[ \frac{m}{\nu}
  K_1 (m \nu )- \frac{m}{\sqrt{r^2+\nu^2}} K_1 (\sqrt{r^2+\nu^2}m)\right]
  .\label{14}\ee

Case \ref{itm:fr} is obtained as $m=0$ limit of (\ref{9})

\be  G^{(1g)}_0(x) = \frac{N_c(N_c^2-1)}{4\pi^2 x^2}, ~~ D_1^{(0)} (x) = \frac{2\alpha_s   (N_c^2-1)}{ \pi x^4 N_c},\label{11}\ee
where the resulting \(V_1\) is also divergent.

The divergence is contained in the perturbative Coulomb part.
To renormalize it, we follow \cite{31,32}
\be V_1^{\rm (pert)} (r)= \frac{8\alpha_s}{3\pi} \int^\infty_0 d\nu\left( \frac{1}{\nu^2}-\frac{1}{\nu^2+r^2}\right) = V_1^{\rm (pert)} (\infty)-\frac{4\alpha_s}{3r}\label{15}\ee
and put $V_1^{\rm (pert)} (\infty)=0$ as it is  accepted in lattice calculations \cite{3,40}. 

As a result, we construct $V_1(r)$ as a sum of two terms: the Coulomb-like  $V_1^{\rm Coul} (r) = - \frac{4\alpha_s}{3 r}$ and $V_1^{(as)}(r)$ given by \eqref{13}
\be  V_1(r) = - \frac{4\alpha_s}{3r} + V_1^{(as)} (r). \label{31b}\ee

\subsection{Nonzero temperature potentials} 

We introduce temperature by means of Matsubara reformulation of the path integral \eqref{5.1} (the ``temporal'' dimension compactification).
  
The $T$-dependence enter $V_1(r, T)$ in two ways. 
The first is the formal substitution of all the infinite ``time''-integration limits with the inverse temperature \(T^{-1}\).

The second is originated by O(4) symmetry breaking. 
Instead of two correlators \(D\), \(D_1\) dependent on distance in the 4d euclidean space, we obtain five correlators $D^E$, $D^E_1$, $D^H$, $D^H_1$, $D^{EH}_1$ dependent on distance in the 3d space and on distance over the compactified coordinate separately.

The correlators \(D^E\) and \(D^H\) yield two string tensions --- confining, color-electric (CE), \(\sigma(T)\) and spatial, color-magnetic (CM), \(\sigma_s(T)\) respectively. 
The string tensions coincide at \(T=0\).
\(\sigma_s\) grows with \(T\).
\(\sigma\) decreases and vanishes \cite{45,46,47,48}.

We simplify the situation using the basic FCM assumption about the locality of field correlators with the correlation length \(\lambda\lesssim 0.2~\)fm \(\sim 1~\text{GeV}^{-1}\). 
This assumption was rigorously justified at \(T=0\) \cite{FCMconf} and was shown to be correct for the magnetic correlators at \(T>0\) \cite{38}. 

Exact consideration would yield corrections of the order of \(e^{-\lambda T}\).
At \(T<400\) MeV, we neglect them --- the locality allows us to treat the correlators as dependent on distance in the 4d compactified space.

Since we introduced the instantaneous potentials in \eqref{5.4}, we should replace \(D\) and \(D_1\) with the corresponding CE correlators.

$V_1(r, T)$ is obtained from \eqref{2} following \cite{31} by the integration limit substitution and replacement of the correlator
  \be V_1(r,T) = \int^{1/T}_0 d\nu (1-\nu T) \int^r_0 \xi d\xi D_1^E (\sqrt{\xi^2+\nu^2}).\label{16}\ee
The same procedure is applied to \(V_D\).
 
Approximating $D_1^E$ with the asymptotic (\ref{6}) we get
\be V_1^{(as)} (\infty, T) = \frac{6\alpha_s \sigma_f}{M_{GP}} \left( 1- \frac{T}{M_{GP}} \left( 1-e^{-\frac{M_{GP}}{T}}\right)\right). \label{26}\ee   
Note that the string tension and the gluelump mass depend on temperature.

For the Coulomb part we obtain
\be V_1^{\rm Coul} (r,T) = - \frac{C_2\alpha_s}{r} \left(1-\frac{2}{\pi} \arctan (rT) - \frac{rT}{\pi} \ln \left( 1+ (rT)^{-2}\right)\right). \label{48}\ee

\section{Polyakov loop in the confined phase and in the transition region}

In this section, we argue on the basis of available lattice data that the free energy in the corresponding definition of \(L (T) = \exp \left(-\frac{F (T)}{T}\right)\) in QCD at \(100\lesssim T\lesssim 400\) MeV is dominated by the heavy-light mass
\be L_i (T) =\exp \left( - \frac{M^{i}_{HL}}{T}\right), ~~ 
    M^f_{HL} = M_{Q\bar q},\,M^{adj}_{HL}=M_{Gg}.\label{3.1}\ee
From the Casimir scaling we deduce that \(L_{adj}\approx L_f^{9/4}\) and, consequently, \(M_{Gg}\approx \frac94 M_{Q\bar q}\).

To obtain the free energy of a static quark (which yields \(L_f\)) we consider a heavy-light meson at various $T$.

At \(T\lesssim 100\) MeV, we content ourselves with the HRG model.

At high \(T\) (in comparison with \(T_c\sim 140\) MeV), in the deconfinement phase, the light antiquark can move arbitrary far away from the static quark.
However, according to the lattice data \cite{54} the CE correlator \(D_1^E\) does not vanish.
Hence, the non-confining interaction \(V_1\) holds, so the quark and antiquark each has free energy of \(V_1(\infty, T)/2\) given by \eqref{26}.
At \(T\gtrsim 400\) MeV the Coulomb interaction $\left \lan \frac{4\alpha_s (T)}{3 r} \right\ran$ of the static quark with other particles of the thermodynamic ensemble determined by the concentration \(n(T) = \frac{\partial P}{\partial T}\) is negligible, so we agree with \eqref{1.5}.

To analyze lower \(T\), we consider the mesons' Hamiltonian. 
Hamiltonian for a meson follows from the path integral representation \eqref{5.1}.
At \(T=0\) for a meson with the orbital momentum \(l=0\) 

\be H_{q\bar q}= \sqrt{\vep_q^2 + m^2_q} + \sqrt{\vep_{\bar q}^2 +m^2_{\bar q}} + V_D^{\rm conf} (r) + V_1^{\rm Coul}(r)+ \Delta V (r) + V_{ss},\label{41}\ee 
where we separated ``saturating'' parts of potentials as \(V_D\equiv V_D^{\rm conf}-V_D^{\rm sat}\), \(V_1\equiv V_1^{\rm Coul}+V_1^{\rm sat}\), introduced the spin-dependent  potential $V_{ss}$ \cite{37}, and denoted \(\Delta V = V_1^{\rm sat} - V_D^{\rm sat}\).
The confining potential is linear $V_D(r) = \sigma r$.

The saturating and confining potentials at \(T>0\) are
\begin{align}   
    &V_D^{\rm conf} (r,T) = 2r\int^{1/T}_0 d\nu (1-\nu T) \int^r_0   d\xi D^E(\sqrt{\xi^2 +\nu^2})\label{46}\\
    &V_D^{\rm sat} (r,T) = 2\int^{1/T}_0 d\nu (1-\nu T) \int^r_0 \xi d\xi D^E(\sqrt{\xi^2 +\nu^2})\label{45}\\
    &V_1^{\rm sat} (r,T) = \int^{1/T}_0 d\nu (1-\nu T) \int^r_0 \xi d\xi D_1^E(\sqrt{\xi^2 +\nu^2})-V_1^{\rm Coul}(r, T).\label{47}
\end{align}
The Coulomb potential was defined in \eqref{48}.
   
The static quark free energy $F_Q(T)$ is produced by all the interactions of the static quark with the  environment via potentials (\ref{45}), (\ref{46}), (\ref{47}), (\ref{48}). 
Correspondingly we represent $F_Q$ as a sum
\be F_Q = M_{Q\bar q} (T) + F_Q^{\rm sat} + F_Q^{\rm Coul},\label{39*}\ee
where $M_{Q\bar q}$ refers to the heavy-light mass, $F_Q^{\rm sat}$  contains the input of (\ref{45}) and (\ref{47}), and $F_Q^{\rm Coul}$ --- the Coulomb interaction. We neglect the spin-spin and spin-orbit interactions.
    
The saturating potentials contribution to the quark free energy is not observed in lattice simulations (see, for example \cite{3,6}) in the confined regime --- linear confining and Coulomb potentials fit the data. 
Here, we show the compensation of the saturating potentials assuming that they are saturated on average at any \(T<T_c\)
(with a more involved calculation we can show that \(\Delta V(r,T)\) is small at any \(r\))
\begin{align}
    F_Q^{\rm sat}(T) &=\frac12 \Delta V (\infty, T)\nonumber \\
    &= \frac12 \int^{1/T}_0 d\nu (1-\nu T) \int^\infty_0 \xi d\xi (D_1 (\sqrt{\xi^2 + \nu^2 }) - 2 D  (\sqrt{\xi^2 + \nu^2 }))\label{48**}
\end{align}
  
To estimate $F^{\rm sat}_Q (T)$, we use $D_1^{as} (x) = A_1 \exp (-M_1 x)$ from (\ref{6}) and  
  \be D^{as}(x) = \frac{g^4 (N^2_c-1)}{2} 0.108 \sigma^2_f e^{-M_2x}= A_2 e^{-M_2 x}\label{48***}\ee
from \cite{41,42}. 
As a result of integration in (\ref{48**}), we obtain at $T=0$ with $M_1=1.4$ GeV and $M_2 = 2$ GeV
  \be F_Q^{\rm sat}(0) = \frac12 \left( \frac{A_1}{M_1^2} - \frac{4A_2}{M^3_2} \right) = \frac12 ( 0.16-0.10)  {~\rm GeV}= 0.03~{\rm GeV}. \label{48****}\ee
  For $T>0$ using  (\ref{26}) we obtain
  \be F_Q^{\rm sat}(T) =    F_Q^{\rm sat}(0)\sqrt{\frac{\sigma(T)}{\sigma(0)}} \left( 1- \frac{T}{M_{GP} (T)} \left( 1- e^{-\frac{M_{GP}(T)}{T}}\right) \right).\label{48*****}\ee
  
From \eqref{48****} and (\ref{48*****}), we deduce that $F_Q^{\rm sat} (T)$ is negligible in the confined regime.
  
As a result, we reduce the static quark free energy (\ref{39*}) to $M_{Q\bar q} (T)$.
The $M_{Q\bar q}$ is dominated by confinement and scales as $\sqrt{\sigma (T)}$. 
The color Coulomb contribution is negative, and it scales according to the same rule in the first approximation, which is sufficient for our present purposes.

To check the result, we need to find $\sigma (T)$ and compare the Polyakov loop \eqref{3.1} with direct lattice measurements.

Direct evidence for the $\sigma (T)$ dependence was found in numerous lattice calculations \cite{45,46,47,48}.
On the other hand, the CE string tension in the massless quarks limit is related to the chiral condensate \cite{49, 50, 51} as $|\lan \bar q q(T)\ran| =const~ (\sigma (T))^{3/2}$.
We introduce dimensionless \(a(T)\) as $\sigma (T) = \sigma (0)  a^2 (T)$, so 

\be |\lan \bar q q\ran (T)| = |\lan \bar q q\ran(0)| a^3 (T)\label{50}\ee

The heavy-light mass $M_{Q\bar q} (T)$ starts at $T=0$ with the minimal eigenvalue of the Dirac equation with the confining and color Coulomb interactions (with \(\alpha_s=0.3\)) $M_{Q\bar q} (0) = M_D \approx 465$ MeV.
To account for the systematic uncertainties of the \(M_{Q\bar q} (0)\) evaluation and the uncertainty of \(\alpha_s\), we estimate the boundaries as \(400~{\rm MeV}<M_{Q\bar q}(0)<585~{\rm MeV}\) \cite{CoulCorr}.
These values are in good agreement with recent lattice calculations \cite{61,62}.

\begin{figure}
    \includegraphics[width=0.85\linewidth]{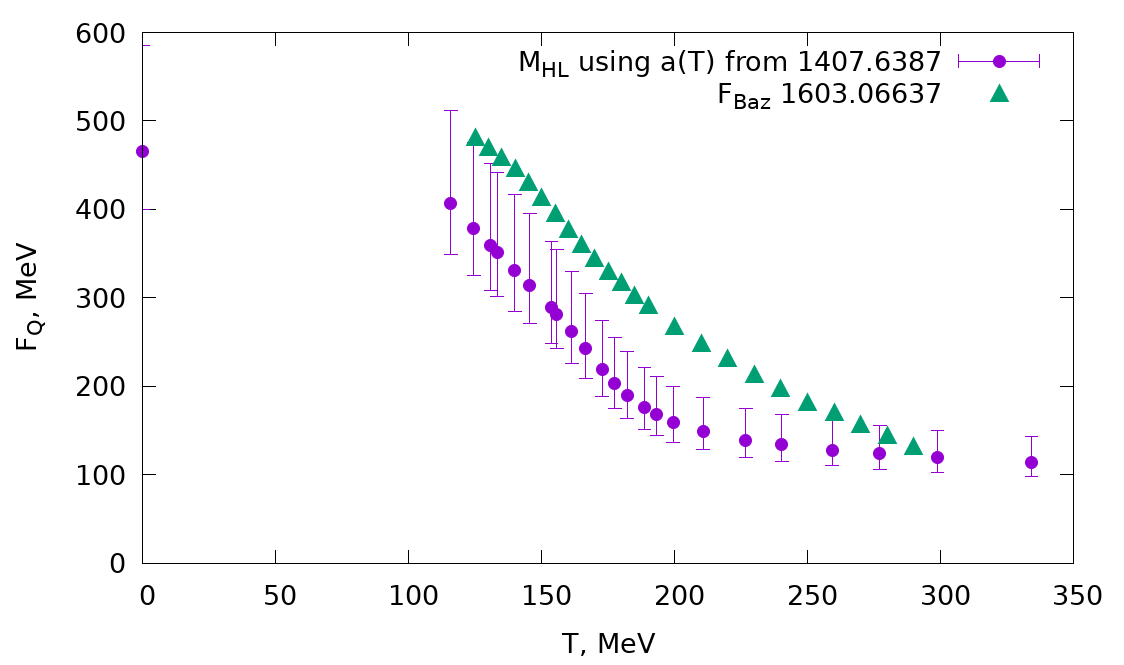} 
    \caption{The free energy \(F_Q\approx M_{HL}\) of a static quark in comparison with the lattice data by Bazavov et al. \cite{59}}\label{f3}
    \includegraphics[width=\linewidth]{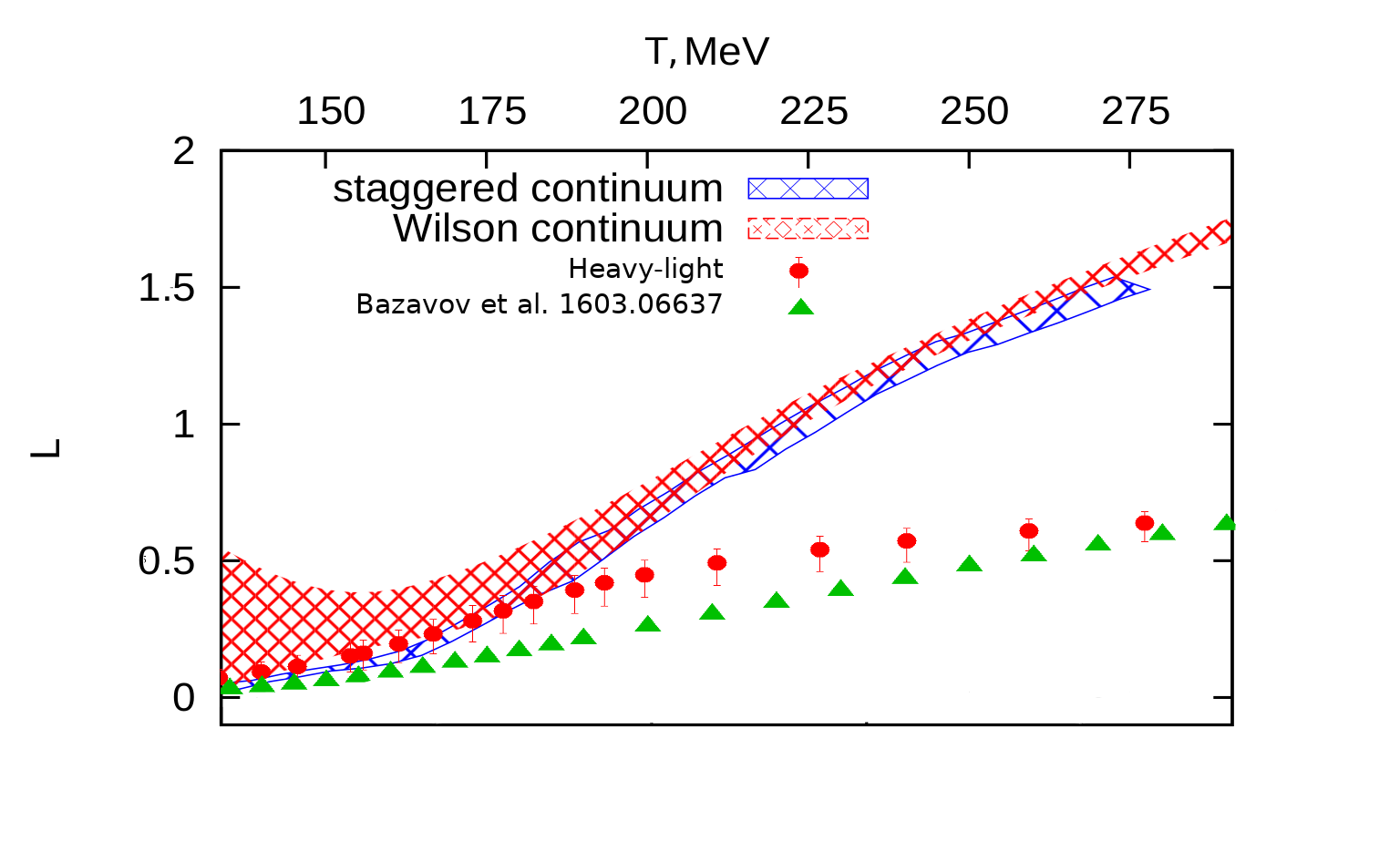} 
    \caption{The fundamental Polyakov loop $L_f(T)$ on $T$ (34) in comparison with the lattice data \cite{56} (upper shaded area) and \cite{59} (the lower line of the points)}\label{f2}
\end{figure}

The resulting behavior of $F_{Q}(T)\approx M_{HL}(T)\approx M_{HL}(0)a(T)$ is shown in Fig. \ref{f3} together with the corresponding behavior of the lattice measured values $F_Q^{Baz} (T)$ \cite{59}. 

To check our result \eqref{3.1} regarding the origin of the Polyakov loop in QCD, in Fig. \ref{f2}, we compare our $L_f(T)$ with lattice $L_f(T)$ calculated by the Wappertal--Budapest group \cite{56,57,58} and by HotQCD group \cite{59}. 
Our dependence falls in the large gap between the lattice results.

\section{Conclusion and discussions}

In this paper, we analyzed the dynamical origin of the Polyakov loop in QCD based on the color-electric (CE) vacuum correlators. 

We found that these correlators produce four types of interactions: confining  $V_D^{\rm conf}$,  saturating $V_D^{\rm sat}$ and $V_1^{\rm sat}$, and color Coulomb interaction $V_1^{\rm Coul}$. 
We provided arguments that all these interactions, except $V_1^{\rm Coul}$, are produced non-perturbatively by the correlators \(D^E_1\) and $D^E$.   

The confining correlator \(D^E\) gives rise to the CE string tension $\sigma =\frac12 \int D^E(x) dx$.
It decreases with temperature, as the lattice data show \cite{45,46,47,48}. 
We used the lattice data \cite{52,53} on chiral condensate $\lan \bar q q (T)\ran$ to find the $T$-dependence of $\sigma (T)$. 

We showed that, to a large extent, $V_D^{\rm sat}$ and $V_1^{\rm sat}$ compensate each other at \(T\lesssim 400\) MeV, and $V^{\rm conf}_D $ and $V^{\rm Coul}$ are left. 
Therefore, the static quark free energy is dominated by the heavy-light hadron mass produced by $V_D^{\rm conf}$ and $V^{\rm Coul}$.

At \(T\lesssim 100\) MeV, HRG satisfactorily describes the QCD Polyakov loop.
At \(T\gtrsim 400\) MeV, a static quark free energy is defined by the remnants of pair interaction \(V_1^{\rm sat}(\infty, T)\).

As a result, $L(T)$ in QCD is determined by the heavy-light mass $M_{Q\bar q}(T)$ at \(100\lesssim T\lesssim 400\) MeV. 
The mass decreases with $T$ as $\sqrt{\sigma(T)}$. 
The resulting $L(T)$ demonstrates behavior 
\footnote{Note that we used the renormalization procedure similar to that of \cite{59}, and  different from \cite{56}.
Hence, the  large difference between the $L(T)$ in the this work and  the data  of \cite{56} in Fig. \ref{f2}, and a  close similarity with \cite{59}.}
similar to the lattice data of \cite{59} supporting the idea that the direct measurement of the Polyakov loop in $n_f=2+1$ QCD on lattice is, to a large extent, a measurement of the heavy-light hadron mass. 

This work  was   done  in the framework of the  scientific project, supported by the Russian Science Foundation grant 16-12-10414.

\end{document}